\documentclass[useAMS,usenatbib]{mn2e}

\usepackage{graphicx}
\usepackage{natbib}

\newcommand\ion[2]{\mbox{#1\,{\sc #2}}}
\newcommand\aap{A\&A}
\newcommand\pasp{PASP}
\newcommand\aj{AJ}
\newcommand\mnras{MNRAS}
\newcommand\apj{ApJ}
\newcommand\aaps{A\&AS}

\title[MP Centauri]{Observational Studies of Early-type Binary Stars: \\ MP Centauri}
\author[Terrell, \it{et al.}]
{Dirk Terrell$^{1}$\thanks{E-mail: terrell@boulder.swri.edu (DT); 
munari@pd.astro.it (UM); tomaz.zwitter@fmf.uni-lj.si (TZ); georgewolf@smsu.edu (GW)},
Ulisse Munari$^{2}$, Toma\v{z} Zwitter$^{3}$ and George Wolf$^{4}$\\
$^{1}$Department of Space Studies, Southwest Research Institute, 1050 Walnut St., Suite 400, Boulder, CO 80302, USA\\
$^{2}$Osservatorio Astronomico di Padova, Sede di Asiago, I-36032 Asiago (VI), Italy\\
$^{3}$Department of Physics, University of Ljubljana, Jadranska 19, 1000 Ljubljana, Slovenia\\
$^{4}$Department of Physics, Astronomy and Materials Science, Missouri State University, Springfield, MO 65804, USA}
\begin{document}

\date{Accepted 2005 March 3. Received 2005 March 3; in original form 2005 March 3}

\pagerange{\pageref{firstpage}--\pageref{lastpage}} \pubyear{2002}

\maketitle

\label{firstpage}

\begin{abstract}
We present photometric and spectroscopic data on the early-type binary MP Centauri. 
The photometric data are analyzed simultaneously with radial velocities to derive preliminary absolute 
dimensions for the binary components. Analysis of the spectra shows that the stars rotate synchronously
and that the line of sight to the system crosses two kinematically sharp and well-separated interstellar
reddening sources. It is shown that MP~Cen consists
 of a B3 primary with $M_{1}=11.4 \pm 0.4 M_{\sun}$, $R_{1}=7.7 \pm 0.1 R_{\sun}$ and a 
lobe-filling B6-B7 secondary with $M_{2}=4.4 \pm 0.2 M_{\sun}$, $R_{2}=6.6 \pm 0.1 R_{\sun}$
\end{abstract}

\begin{keywords}
binaries: eclipsing -- binaries: spectroscopic -- stars: individual (MP Cen) 
 -- stars: evolved.
\end{keywords}

\section{Introduction}

MP Centauri (HD 308976) is an eclipsing binary, listed in the HD catalog as having a B3 spectral type. 
Its orbital period of 2.99 days makes it a very difficult target to obtain complete 
light curves from a single location in one observing season. It is a member a rare
group: massive eclipsing binaries with short orbital periods. It lies in the galactic plane
($b=0^{\circ}.08$, $l=295^{\circ}.01$) and its eclipsing nature allows for the determination of
its absolute parameters and distance, as well as the reddening along the line of sight. We have 
obtained $uvby$ photometry over several years and the system was also observed in $V$ and $I_C$ 
by the All Sky Automated Survey (\citealt{poj02}; hereafter ASAS).

As part of a program on 
early-type overcontact and near-contact binaries, we observed MP Cen with the ESO 
La Silla 2.2m telescope and the FEROS spectrograph at high resolution (R=48,000) in 
February, 2003. Combining radial velocities from these spectra with the $uvby$ photometry 
and $V$ and $I_C$ data from ASAS, we present the first comprehensive study of this evolved binary.

\section{Observations}
\subsection{Photometry}

GW observed MP Cen in 1982 and 1989 using the 24-inch (0.6 meter) telescope and single
channel photometers equipped with $uvby$ filters at the Mt. John Observatory at Lake Tekapo,
New Zealand. The comparison star used was HD 102139, for which the SIMBAD database gives a B4 III spectral 
type. A 17 arcsec aperture was used and offset slightly to avoid contamination from nearby 
companions.

The ASAS project has also obtained $V$ and $I_C$ CCD photometry of MP Cen. 
Because the goal is to survey large areas of the sky, the ASAS setup is not optimal for
photometry of stars in crowded fields like MP Cen. However, MP Cen is the brightest object 
in the measuring aperture and a proper treatment of the third light ($l_3$) contamination
makes it possible to use the ASAS data. The scatter in all passbands was about 2\%.

\subsection{Spectroscopy}

Seven ESO 2.2m + FEROS spectra have been
secured, in two groups of three spectra each on 2003 Feb 23 and 24 near quadratures and a single spectrum
on 2004 Feb 25 in primary eclipse (Table 1). The wavelength range of the
spectra extends from 3900~\AA\ to 9200~\AA\ with $R=48,000$ and a 1200 second
exposure time. The S/N is evaluated 
around 5850~\AA\ and the orbital phase is computed with the ephemeris given in section \ref{analysis}.

\begin{table*}
 \begin{minipage}{140mm}
 \label{speclist}
  \caption{ESO 2.2m + FEROS Spectra of MP Cen.}
  \begin{tabular}{ccccc}
  \hline
  Spectrum Number & Date & UT of mid-exposure & S/N  & Orbital Phase\\
  \hline
  1440, 1441, 1442 & 2003 Feb 23 & 05:40 &  75 & 0.28 \\   
  1552, 1553, 1554 & 2003 Feb 24 & 09:15 &  77 & 0.66 \\
  1632             & 2003 Feb 25 & 06:48 &  85 & 0.96 \\
  \hline
  \end{tabular}
 \end{minipage}
\end{table*}

Our estimate of the spectral type of the primary, based on the relative intensities of 
\ion{Mg}{ii}~4481, \ion{He}{i}~4471, \ion{C}{ii}~4267, \ion{He}{i}~4388, \ion{He}{i}~4009
and \ion{He}{i}~4026, is B3 with an estimated error of one spectral subclass. Because of 
the relative faintness of the secondary and its lines, it is more difficult to estimate 
its spectral type but a value in the range B6 to B7 seems reasonable, based on the ratio 
of \ion{He}{i}~4471 to \ion{Mg}{ii}~4481, and is consistent with our light curve solution. 

\subsubsection{Rotational velocity}

To derive the rotational velocities we used the relation
\begin{equation}
\label{reddening}
V_{\rm rot} = 42.42 \times HIW\ -\ 35  {\rm km~sec^{-1}}
\end{equation}
\noindent
calibrated by \cite{mun99} on the width at half maximum (HIW) of
\ion{He}{i}~5876~\AA\ in high resolution spectra of O and B stars. The deconvolution
of the \ion{He}{i}~5876~\AA\ line profile in the MP Cen quadrature spectra gives a HIW of
4.12~\AA\ for the primary, and 3.18~\AA\ for the secondary, which corresponds to 140 and 
100~km~sec$^{-1}$ respectively. 

The Munari and Tomasella relation was calibrated using single, 
and thus axially symmetric, stars. In previous work on the overcontact binary TU~Mus \citep{ter03}, 
we found that it was necessary to correct the HIW for the distorted shapes of the two stars.
The MP~Cen primary is reasonably spherical so no correction is necessary, but the secondary is
highly distorted so, following the procedure outlined in \cite{ter03}, we find that the 
corrected rotational velocity of the MP~Cen secondary is 95~km~sec$^{-1}$. The profiles, however, 
are not indicative of purely rotational broadening, with atmospheric effects and circumstellar
activity probably distorting the profile. In the absence of more sophisticated modeling, 
we see no evidence of asynchronous rotation and thus assume synchronism in the rest of our 
analysis.

\subsubsection{Reddening}

\begin{figure}
 \includegraphics[width=84mm]{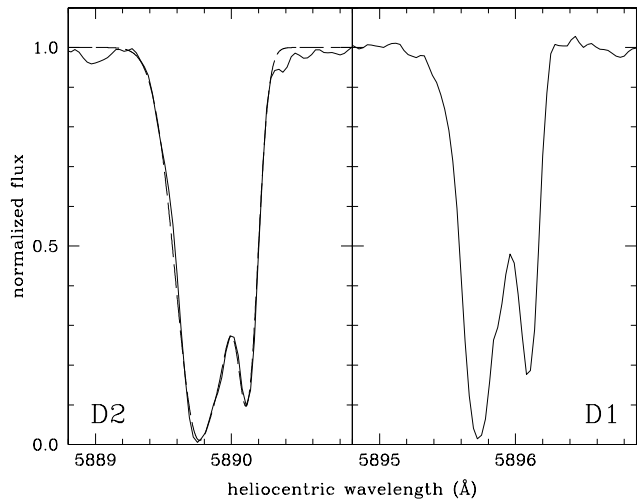}
 \caption{The interstellar lines of \ion{Na}{i}, averaged from seven spectra, with the two-component 
Gaussian fit to the D2 line used in the reddening determination. The similar shape of the D1 line
shows that the shape of D2 is not an artifact of noise or atmospheric lines.} 
 \label{naiplot}
\end{figure}

\cite{mun97} demonstrated a method of using the interstellar \ion{Na}{i} and \ion{K}{i} lines to 
estimate the reddening affecting stars, independent of the knowledge of the intrinsic
colors. The profile of \ion{Na}{i} D1 and D2 lines in our FEROS spectra of MP~Cen
are shown in Figure \ref{naiplot} and are clearly composed of at least two 
components.  The 7 individual spectra give very consistent results and indicate a
value of $E(B-V)=0.30$. The two-Gaussian {\em mean} fit is overplotted on the D2 line 
in Figure \ref{naiplot} to show the accuracy of the fit. 

The heliocentric radial velocities of the two  components are $6.23 \pm 0.13$ km sec$^{-1}$
and $23.57 \pm 0.08$ km sec$^{-1}$. The accuracy of these velocities is not fictitious 
because nearby telluric absorptions show the same wavelength stability from one spectrum to the
other, and it matches the expected high FEROS spectrograph stability, suitable for extra-solar planet searches.
The line of sight to MP~Cen therefore crosses two kinematically very sharp and well separated 
sources of reddening of purely interstellar origin since neither of the components shows a 
radial velocity change with orbital phase nor shares the systemic velocity of the binary.

The Tycho $B_T - V_T = 0.13 \pm 0.04$ corresponds, according to \cite{b00}, to a Johnson
$(B-V) = 0.11$. We denote an intrinsic (\it i.e.\rm, unreddened) color by $(B-V)^\prime$ and an 
observed color by $(B-V)$ with subscripts $1$, $2$ and $C$ denoting the primary 
component, the secondary component and the composite value for the binary respectively.
With an assumed intrinsic color
for the primary, $(B-V){}_1^\prime$, and the magnitude differences between the two components
in each filter, $\Delta m_B$ and $\Delta m_V$, from our light curve solution,
we can compute the intrinsic composite color of the binary, $(B-V){}_C^\prime$, as
\begin{displaymath}
    (B-V){}_C^\prime=(B-V){}_1^\prime - 2.5\; \log{
                               \left( \frac{1+10^{(-0.4\times\Delta m_B)}}
                                    {1+10^{(-0.4\times\Delta m_V)}} \right) 
                              }
\end{displaymath} 
Assuming a value of $(B-V)^\prime = -0.22$ for the B3 primary \citep{pjf96} and the $\Delta m_B$ and
$\Delta m_V$ values of 1.16 and 1.12 respectively from the light curve solution, we find
$(B-V){}_C^\prime = -0.21$. This result, combined with the observed $(B-V)_C$, yields a reddening 
of $E(B-V)=0.32$, in excellent agreement with the results from analysis of the interstellar 
\ion{Na}{i} and \ion{K}{i} lines.

\noindent

\subsubsection{Radial velocities}

\begin{figure}
 \includegraphics[width=84mm]{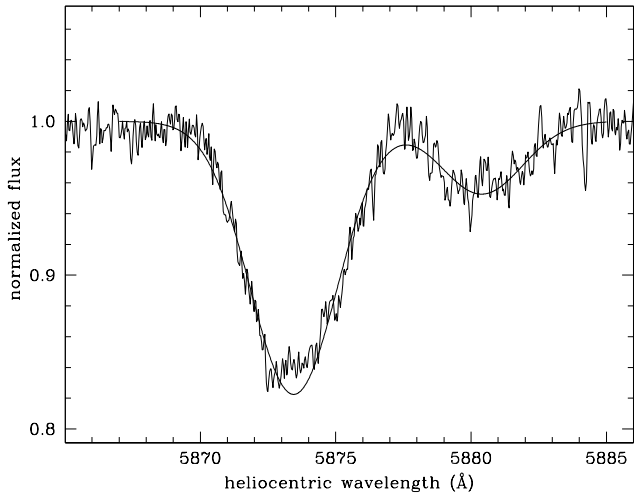}
 \caption{The \ion{He}{i} 5876 profile of MP~Cen for the averaged Feb 24 spectra. The overplotted 
curve is the fit of two gaussians at the radial velocities of the two components given in Table 2.}
 \label{lineplot}
\end{figure}

The radial velocities of the MP~Cen components have been measured on the two groups of spectra
in Table 1
obtained close to quadrature phases. The spectrum taken at 
orbital phase 0.96 is very close to conjunction and thus has insufficient velocity 
separation between the  two components to obtain reliable radial velocities.
Line splitting at quadrature is wide and allows an easy and firm determination 
of the radial velocity, as shown for \ion{He}{i} 5876 in Figure \ref{lineplot}. 
The radial velocities have been obtained by Gaussian de-convolution of the profiles into two
components, which provide a good overall fit to the observed line
profile. The measured lines are \ion{He}{i}~4016, 4388, 4471, 5876, 6678, 7065 and 
half-weight contributions from H$\gamma$, H$\beta$ and H$\alpha$. The means
of the radial velocities (and associated standard errors) derived from these
lines are given in Table 2.

\begin{table}
 \label{rvtable}
  \caption{Heliocentric radial velocities of MP~Cen.}
  \begin{tabular}{cccccccc}
   \hline
    Spectra & Phase
    & Primary (km sec$^{-1}$) 
    & Secondary (km sec$^{-1}$)\\
   \hline
    1440-1442 & 0.28 &$-97.4\pm2.7$  &$+256.0\pm4.8$ \\ 
    1552-1554 & 0.66 &$+88.6\pm2.3$  &$-216.9\pm3.3$ \\ 
   \hline
  \end{tabular}
\end{table}

\section{Data Analysis}
\label{analysis}

We performed a simultaneous analysis of the ASAS photometry, our $uvby$ photometry and our radial 
velocities with the 2003 version of the Wilson-Devinney (\citealt{wd71}; \citealt{rew79}; \citealt{rew90};
hereafter, WD) program. As we had limited telescope
time available to us for spectroscopy, we could not obtain as many radial velocities as we 
would have liked and, therefore, we present our solution as preliminary until more extensive spectroscopic 
observations can be made. However, our spectra are of high quality and resolution, giving us 
confidence that our results are reliable and certainly better than a solution based solely on
the photometry. We have shown previously that our velocities for TU Mus \citep{ter03}, obtained 
in the same observing run as the MP Cen spectra, were in excellent agreement with earlier data 
from \cite{ag75}. 

We also performed tests of our solution results by generating 1000 sets of 
synthetic radial velocities with Gaussian errors based on our four velocities and their estimated errors.
We solved each set of four velocities and found that the average and standard deviation of 
each parameter was consistent with the results of our full solution. In particular, the average value of 
the semi-major axis of the relative orbit, of critical importance in determining the absolute dimensions
of the binary,  was $21.9 \pm 0.2 R_{\sun}$, matching exactly the result from the full solution.

WD now uses \cite{k93} stellar 
atmospheres to model the radiation from the stars. WD can also use either binary orbital phase
or time as the independent variable \citep{wt98}. In order to investigate possible changes in the 
orbital period, we used time as the independent variable and adjusted the orbital period ($P$), 
its first time derivative ($\dot{P}$), and the reference epoch ($HJD_0$). We began
our light curve fitting experiments in WD's mode 2, appropriate for detached binaries \citep{lw77},
but found that a semidetached configuration with the secondary filling the lobe was required 
to fit the observations so we continued our analysis in WD's mode 5.

Other parameters adjusted in the simultaneous solution were the semi-major axis of the relative orbit
($a$), the binary centre of mass radial velocity ($V_{\gamma}$), orbital inclination ($i$),
secondary mean effective temperature ($T_2$), modified surface potential of the 
primary ($\Omega_1$), mass ratio ($q$), and the bandpass-specific luminosity of the primary 
($L_1$). Certain parameters, such as the
bolometric albedos and gravity brightening exponents, were held fixed at their expected theoretical 
values. The logarithmic limb darkening law was used with coefficients from \cite{wvh93}. The 
mean effective temperature of the primary was set to 18,750 K based on the B3 spectral 
type \citep*{b98}. Data set weights were determined by the scatter of the observations. 

Given the large apertures used in the ASAS photometry and the moderately crowded field 
around MP Cen, we adjusted third light and found statistically significant values 
for the ASAS data but not our $uvby$ photometry. Images of the MP Cen field show two 
companions to the northeast about 13" away. These stars fit within the ASAS aperture but
were avoided in our $uvby$ photometry. 

\begin{figure}
 \includegraphics[width=84mm]{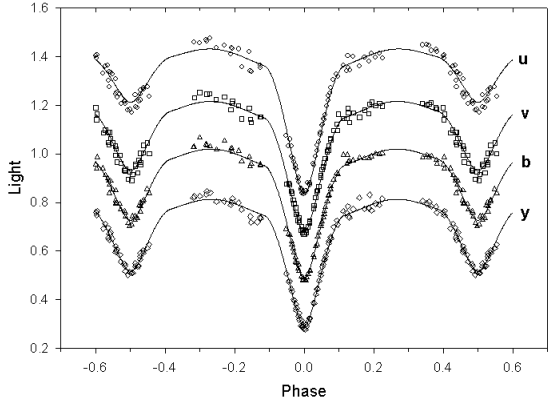}
 \caption{The Str\"{o}mgren photometry and computed light curves.}
 \label{uvbyfits}
\end{figure}

\begin{figure}
\includegraphics[width=84mm]{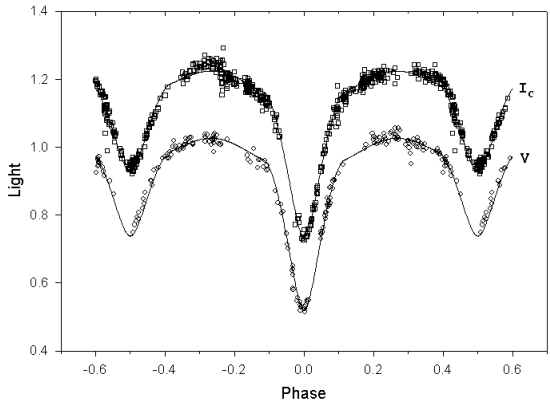}
\caption{The ASAS $V$ and $I_C$ photometry and computed light curves.}
\label{asasfits}
\end{figure}

\begin{figure}
\includegraphics[width=84mm]{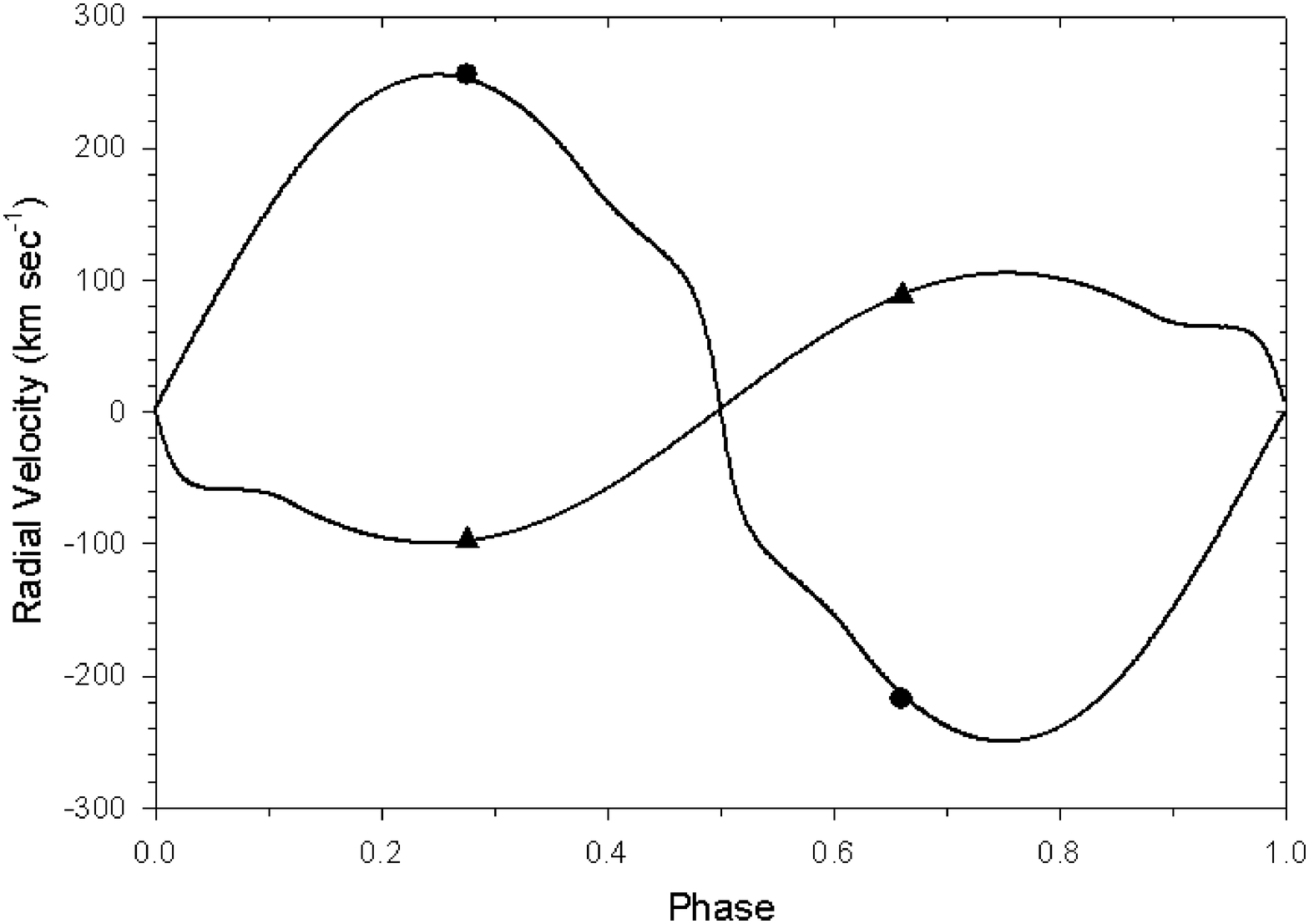}
\caption{The observed radial velocities and computed curves.}
\label{rvfits}
\end{figure}

Table \ref{parameters} shows the results of 
the simultaneous solution and Figures \ref{uvbyfits} and \ref{asasfits} show the 
fits to the Str\"{o}mgren and ASAS data respectively. Figure \ref{rvfits} shows the fits to 
the radial velocities. The light curves show some disturbances, particularly at the maximum preceding 
primary eclipse, as expected if a mass transfer stream from the secondary is impacting
the primary.

\begin{table}
  \caption{Parameters of MP Cen}
  \label{parameters}
  \begin{tabular}{cc}
  \hline
   Parameter & Value\\
  \hline
  $a$ & $21.9 \pm 0.2 R_{\sun}$  \\
  $V_{\gamma}$ & $2.7 \pm 1.7$ km sec$^{-1}$\\
  $i$ & $82^{\circ}.2 \pm 0^{\circ}.2$  \\
  $T_{1}$ & $18,750$ K \\
  $T_{2}$ & $12,390 \pm 50$ K\\
  $\Omega_{1}$ & $3.29 \pm 0.03$ \\
  $q$ &  $0.390 \pm 0.008$ \\
  $HJD_{0}$ &  $2447627.9453 \pm 0.0006$ \\
  $P$ & $2.993456 \pm 0.000001$ days \\
  $\dot{P}$ & $8.7 \times 10^{-9} \pm 6.1 \times 10^{-10}$ \\
  $L_{1}/(L_{1}+L_{2})_u$ & $ 0.83 \pm 0.01$  \\
  $L_{1}/(L_{1}+L_{2})_v$ & $ 0.76 \pm 0.01$  \\
  $L_{1}/(L_{1}+L_{2})_b$ & $ 0.75 \pm 0.01$  \\
  $L_{1}/(L_{1}+L_{2})_y$ & $ 0.74 \pm 0.01$  \\
  $L_{1}/(L_{1}+L_{2})_V$ & $ 0.74 \pm 0.01$  \\
  $L_{1}/(L_{1}+L_{2})_I$ & $ 0.72 \pm 0.01$  \\
  $l_{3} (V)$ & $ 0.04 \pm 0.02$  \\
  $l_{3} (I)$ & $ 0.05 \pm 0.02$  \\
  $l_{3} (uvby)$ & $ 0.00 \pm 0.02$  \\
  $R_{1}$ & $7.7 \pm 0.1$ R$_{\sun}$ \\
  $R_{2}$ & $6.6 \pm 0.1$ R$_{\sun}$ \\
  $M_{1}$ & $11.4 \pm 0.4$ M$_{\sun}$ \\
  $M_{2}$ & $4.4 \pm 0.2$ M$_{\sun}$ \\
  $log~L_{1}/L_{\sun}$ & $3.8 \pm 0.2$  \\
  $log~L_{2}/L_{\sun}$ & $3.0 \pm 0.2$ \\
  $log~g_{1}$ & $3.7 \pm 0.2$  \\
  $log~g_{2}$ & $3.4 \pm 0.2$ \\
  \hline
  \end{tabular}

\medskip
Quoted errors are formal 1-$\sigma$ errors from the solution.
$l_3$ values are in units of total system light at phase 0.25.
Luminosity errors are estimates based on an uncertainty of 2000 K
in the effective temperature of the primary. The error contributions to the luminosities due to the
errors in the radii are more than an order of magnitude smaller. The $log~g$ values are in CGS units.
\end{table}

The ASAS data show that the system has $V=9.95$ and with the third light correction from our light
curve solution, $V=9.99$ for the binary at maximum brightness. Correcting this value for the reddening 
(assuming a ratio of total to selective absorption of $R_V=3.1$ and our $E_{B-V}=0.30$), we find that
the apparent V magnitude of the binary is 9.08. Using the $V$ luminosity ratio $L_2/L_1$=0.36 from 
our light curve solution, this transforms into Johnson V magnitudes of $V_1$=9.41 and $V_2$=10.53.
With our log luminosity of $3.8 L_{\sun}$ for the primary, the distance to the primary 
amounts to 3.0 kpc using a bolometric correction of -1.75 from \cite{pjf96}. The same calculation
for the secondary results in the same distance. We therefore adopt 3.0 kpc
as the distance to MP~Cen, placing it on the far side of the Sagittarius arm.

MP~Cen lies in the galactic plane and we can
compare our estimate of the reddening with the \cite{nek80} extinction maps. 
MP Cen lies at the boundary of four cells (294/-1, 296/-1, 297/0 and 292/2). 
Given the low resolution of the cell boundaries, we cannot unambiguousily assign MP Cen to a 
particular cell. Our value of $E_{B-V}$=0.30 leads to $A_V = 0.93$. Looking at the 
individual cells, we find
\begin{itemize}
  \item 294/-1  distance greater than 1 kpc
  \item 296/-1  the distance could be as high as 4 kpc
  \item 297/0   distances from 1 to 4 kpc supported
  \item 292/2   distances from 0.5 to 4 kpc supported
\end{itemize}
Thus, the \cite{nek80} maps do not provide a conclusive comparison with our derived distance,
but their Figure 9 supports the idea that the line of sight to MP Cen goes through
clouds A and G and just misses clouds L, M and R.

\section{Conclusions}

We have presented preliminary values for the parameters of MP~Cen based on photometry
and limited spectroscopic data.
The masses and radii of the stars, along with the semidetached configuration,
indicate that the system is evolved. The radii, $R_1=7.7 R_{\sun}$ and $R_2=6.6 R_{\sun}$, 
are consistent with a giant or sub-giant classification for both stars for the derived masses
of $M_1 = 11.4~M_{\sun}$ and $M_2 = 4.4M_{\sun}$. The $\dot{P}$ we have measured implies 
a conservative mass transfer rate of $1.1 \times 10^{-7}~M_{\sun}~yr^{-1}$, making it a fairly active 
system. Light curve variations from season to season, most likely arising from this mass 
transfer activity,  are apparent in the photometric data.
We also see emission features in our spectra but the limited phase coverage of the spectra
makes any discussion of the circumstellar environment little more than speculation at this point.
More extensive spectroscopy and polarimetry will be required to fully map the circumstellar 
environment of the system.

\section*{Acknowledgments}

This research has made use of the SIMBAD database, operated at CDS, 
Strasbourg, France.

\bsp

\label{lastpage}


\begin{thebibliography}{}
\bibitem[\protect\citeauthoryear{Andersen \& Gr\o nbech}{1975}]{ag75} Andersen, J. \& Gr\o nbech, B.
    1975, \aap, 45, 107
\bibitem[\protect\citeauthoryear{Bessell, Castelli \& Plez}{Bessell, et al.}{1998}]{b98} Bessell, M. S., Castelli, F , B. 
   \& Plez 1998, \aap, 333, 231
\bibitem[\protect\citeauthoryear{Bessell}{2000}]{b00} Bessell, M. S. 2000, \pasp, 112, 961
\bibitem[\protect\citeauthoryear{Cox}{2000}]{cox00} Cox, A.N. 2000, Allen's Astrophysical Quantities, 
   (American Institute of Physics Press)
\bibitem[\protect\citeauthoryear{Flower}{1996}]{pjf96} Flower, P. J. 1996, \apj, 469, 355
\bibitem[\protect\citeauthoryear{Kurucz}{1993}]{k93} Kurucz, R. L. 1993, in Light Curve Modeling of Eclipsing Binary Stars,
    , ed. E. F. Milone (New York: Springer-Verlag), 93
\bibitem[\protect\citeauthoryear{Leung \& Wilson}{1977}]{lw77} Leung, K.-C. \& Wilson, R. E. 1977, \apj, 211, 853
\bibitem[\protect\citeauthoryear{Munari \& Tomasella}{1999}]{mun99} Munari, U. \& Tomasella, L. 1999, \aap, 343, 806
\bibitem[\protect\citeauthoryear{Munari \& Zwitter}{1997}]{mun97} Munari, U. \& Zwitter, T. 1997, \aap, 318, 269
\bibitem[\protect\citeauthoryear{Neckel \& Klare}{1980}]{nek80} Neckel, T. \& Klare, G. 1980, \aaps, 42, 251
\bibitem[\protect\citeauthoryear{Pojmanski}{2002}]{poj02} Pojmanski, G. 2002, Acta Astron., 52,397
\bibitem[\protect\citeauthoryear{Terrell, et al.}{2003}]{ter03} Terrell, D., Munari, U., Zwitter, T. \& Nelson, R. H. 2003, 
    \aj, 126, 2988
\bibitem[\protect\citeauthoryear{Van Hamme}{1993}]{wvh93} Van Hamme, W. 1993, \aj, 106, 2096
\bibitem[\protect\citeauthoryear{Wilson}{1979}]{rew79} Wilson, R. E. 1979, \apj, 234, 1054
\bibitem[\protect\citeauthoryear{Wilson}{1990}]{rew90} Wilson, R. E. 1990, \apj, 356, 613
\bibitem[\protect\citeauthoryear{Wilson}{2001}]{rew01} Wilson, R. E. 2001, Inf. Bull. Variable Stars, No. 5076
\bibitem[\protect\citeauthoryear{Wilson \& Devinney}{1971}]{wd71} Wilson, R. E \& Devinney, E. J.
    1971, \apj, 166, 605 (WD)
\bibitem[\protect\citeauthoryear{Wilson \& Terrell}{1998}]{wt98} Wilson, R. E \& Terrell, D.
    1998, \mnras, 296, 33
\end{thebibliography}
\end{document}